\documentclass[a4paper,11pt]{article}
\pdfoutput=1 

\usepackage{jcappub} 

\usepackage[T1]{fontenc} 
\usepackage{graphicx}	
\usepackage{amsmath}	
\usepackage{amssymb}	
\usepackage{breqn}
\title{Impact of Neutrino Properties and Dark Matter on the Primordial Lithium Production}

\author[a]{Tahani R. Makki}
\author[b]{Mounib F. El Eid}
\author[c]{Grant J. Mathews}

\affiliation[a]{Department of Physics, American University of Beirut, Lebanon}
\affiliation[b]{Department of Physics, American University of Beirut, Lebanon}
\affiliation[c]{Department of Physics, Center for Astrophysics, University of Notre Dame,USA}

\emailAdd{trm03@mail.aub.edu}
\emailAdd{meid@aub.edu.lb}
\emailAdd{gmathews@nd.edu}

\abstract{The light elements up to ${}^7$Li were produced by the Standard Big Bang nucleosynthesis (SBBN) in the early universe assuming standard conditions. All observed primordial abundances of these light elements match very well the predicted ones by this SBBN except for ${}^7$Li which seems to be overproduced. It is rather challenging to resolve this discrepancy owing to the diverse possibilities affecting the abundance of lithium. In the present work we focus on non-standard possible solutions such as variation of chemical potential and the neutrino temperatures. In addition, the effect of dark matter is also analysed. We find that including these non-standard assumptions helps to reduce the abundance of lithium as predicted by SBBN. We suggest that this could be a possible step toward resolving the lithium problem.}

\begin{document}
\maketitle
\flushbottom

\section{Introduction}
The lithium problem is difficult to resolve. The reason is that not only the nuclear physics involved in its production should be considered, but also astrophysical and non-standard cosmological aspects. The main concern of this work is to investigate the role of neutrinos and  dark matter on the production of light elements during Big Bang Nucleosynthesis (BBN) especially lithium. It is emphasized that these non-standard scenarios require observational constraints on light elements and updated parameters from the analysis of the Cosmic Microwave Background (CMB) in particular the baryon-to -photon ration $\eta$ and the relativistic degrees of freedom $N_{eff}$. This paper is outlined as follows: In section 2, an overview of the lithium problem is presented. In section 3, we discuss the limits on $N_{eff}$ from cosmology and astrophysics, this will be an important parameter in constraining a non-standard scenario. In sections 4 and 5 we investigate whether neutrinos and axions are possible candidates for dark matter. Section 6 will deal with a model of varying neutrino temperatures as treated in more detail in \cite{Galvez and Sherrer 2017}. Since this model alone does not help to reduce the lithium abundance, we extend this calculation by including neutrino chemical potentials. This is presented in section 7. The results lead to a decrease of lithium at the expenses of large $N_{eff}$ which is not supported by recent CMB observations. Therefore, in section 8, by including the effect of photon cooling with axion dark matter, we can achieve a reduction of lithium with $N_{eff}$ compatible with recent CMB observations but still higher than that inferred from Planck analysis \cite{Plank Collaboration 2016}. To satisfy the requirements on $N_{eff}$ as predicted by Planck, we have treated the effect of a dark fluid \cite{Arbey A.and Mahmoudi F. 2008} along with non-standard neutrino properties. Concluding remarks are given in section 10.
\section{The Lithium Problem}
The observed light-element abundances are in agreement with those predicted by the SBBN except for ${}^7$Li as indicated by its observed abundance in metal-poor halo stars. In addition, the extension of observations to very low metallicity below $[Fe/H]=-3$ are inconsistent with the expected "Spite Plateau" \cite{Spite F. and Spite M. 1982} requiring a constant abundance of lithium as metallicity decreases. Up to now, resolving this problem has not been achieved on purely astrophysical and nuclear physics grounds. Therefore, it is worth focusing on non-standard scenarios including non-standard neutrino properties along with dark matter effects. The goal of these non-standard scenarios is to decrease the abundance of lithium below that predicted by SBBN while taking into account the observational constraint on other light elements. These conservative constraints are adopted from many observations as follows \cite{Fernandez2018, Balashev S. A. 2016, cyburt2016, Plank Collaboration 2016}:
\begin{equation}
\begin{split}
0.228\leq Yp\leq 0.260, \thickspace \thickspace \thickspace
2.58\times{10}^{-5}\le\frac{D}{H}\le 3.75\times{10}^{-5}\\
\frac{{}^7{Li}}{H}\le 2.8\times{10}^{-10}
\label{Eq.0}
\end{split}
\end{equation}
The limit on lithium as above is considered to present the primordial lithium abundance so that achieving this limit by non-standard treatment is considered to be sufficient to match the observations. Any remaining difference can be explained by stellar processing.
\section{Neutrinos and the effective degrees of freedom}
One possible extension of the SBBN is to allow the number of neutrinos to be different from three. However, the effective number of relativistic species $N_{eff}$ is not allowed to vary freely due to its effect on both the CMB and SBBN predictions.
The combination of seven-year WMAP data with Baryon Acoustic Oscillations in the distribution of galaxies and the Hubble parameter $H_{0}$ leads to  $N_{eff}=4.43_{-0.88}^{+0.86}$ \cite{Komatsu 2011}. However, the nine-year WMAP put a more stringent limit on the effective number of relativistic species to be $N_{eff}=3.84\pm0.40$. A stronger limit is given by Planck collaboration \cite{Plank Collaboration 2016} such as $N_{eff}=3.15\pm0.23$.
While limits on $N_{eff}$ from SBBN were intensively investigated \cite{Fields2006,Steigman2012}, limits from astrophysics and cosmology are also important. We mention here the limit given by \cite{Leistedt2014} where constraints on the number of effective relativistic degrees of freedom are deduced form the CMB lensing, baryon acoustic oscillations, and galaxy clustering data to be $N_{eff}<3.8$. Limits on  $N_{eff}$ are still a matter of debate and we still need more accurate determination of $N_{eff}$. So based on the ranges above, in what follows we will adopt $2<N_{eff}<5$ as reasonable limits when considering non-standard scenarios.

\section{Are Neutrinos candidate of dark matter}
While we know several aspects of dark matter such as energy density and distribution, we don't have much information about its identity and production mechanism. Since baryons are not contributing to dark matter, one can think of neutrinos or their decay products as possible candidates due to the fact that neutrinos now have mass. Sterile neutrino, which is a hypothetical new generation of neutrinos other than the three active species, can be produced non-thermally by active-sterile mixing and particle decays \cite{Merle2017}. The role of sterile neutrinos in cosmology strongly depends on the magnitude of of their mass so that a sterile neutrino with mass in Kev could be a viable candidate of dark matter \cite{Boyarsky2006}. If this would be the case, this will be detectable in the extragalactic x-rays due to its radiative decay channel \cite{Dolgov2002}. Constraints on the properties of a dark matter sterile neutrino (sterile neutrino mass and mixing parameter $\theta$ between active and sterile neutrinos)are given in \cite{Boyarsky2006} where Milky Way halo and halo of dwarf galaxies are the best objects for the search of dark matter with radiative decay channel.\\
A new investigation for the possibility of detection of sterile neutrinos of mass 50 kev in dark matter searches is given by \cite{paraskevi2018} which are confronted by two problems: the expected event amount of energy to be received by the detector and the expected event rate. Although sterile neutrinos are heavy, they cannot be detected by standard dark matter experiments. For these reasons electron neutrino scattering is considered by using systems with very small electron binding in order to have high event rate. In addition, considering the options of nuclear physics (the absorption of an antineutrino on electron capturing nuclear system) and atomic physics (possibility of spin induced excitations) can be useful in detecting sterile neutrino dark matter \cite{paraskevi2018}. Therefore, since explaining dark matter on the basis of standard model is not likely, many searches with astrophysical and laboratory experiments are used to check the possibility of a sterile neutrino dark matter candidate \cite{Adhikari2017}.
\section{Axion as dark matter}
Axions produced during the QCD phase transition could be possible candidate of cold dark matter (CDM) where its density is well determined by WMAP and Planck missions. These axions dark matter have average momentum of order of the Hubble expansion rate and they satisfy the CDM density if they would have a mass of order of $10^{-5}$ev/c$^{2}$. With this mass range, axions interact weakly through all forces except gravity which make them one of the promising candidates for CDM \cite{Erken2012}.\\
An investigation for interaction of axion-like dark matter with gluons by searching for a time oscillating neutron is performed by \cite{Abel2017} as well as the effect of axion-wind spin-precession in order to search for the interaction of axion-like dark matter with nucleons. However, this was not promising for placing limits on such interactions rather this have led to improve upon existing astrophysical limits on the axion-gluon coupling and existing laboratory limits on the axion-nucleon coupling.\\
A question arise now: what is the effect of axion dark matter on BBN? An example of such effect was investigated by \cite{Erken2012} where photon cooling through the gravitational field of cold axions after BBN had a significant effect on final element abundances especially on lithium. Photon cooling between the end of BBN and decoupling implies that the baryon to photon ratio $\eta_{BBN}$ during BBN is different from  $\eta_{WMAP}$ which is the one measured by WMAP . More explicitly, the photon cooling implies the energy conservation \cite{Erken2012}:
\begin{equation}
\rho_{i,\gamma}= \rho_{f,\gamma}+ \rho_{f,a}
\label{Eq.00}
\end{equation}
where $\rho_{i,\gamma}, \rho_{f,\gamma}$ are the initial/final energy density of photons and $\rho_{f,a}$ is the final energy density of axions respectively (it is assumed that the energy density of the initial axions and baryons is negligible).
It follows from Eq. (~\ref{Eq.00}) that after reaching the thermal equilibrium, the initial and final photon temperatures are related through:
\begin{equation}
T_{f} = (2/3)^{1/4}T_{i}
\label{Eq.1}
\end{equation}
Since the number density of photons is proportional to $T^{3}$, it is straightforward that:
\begin{equation}
\eta_{BBN} =(\frac{2}{3})^{3/4}\eta_{WMAP}
\label{Eq.2}
\end{equation}
This model is treated by \cite{Erken2012} in order to see if it solves the lithium problem. Although this model suppresses the conflict between BBN predictions and observations of lithium, this have led to an overproduction of deuterium and an increase in $N_{eff}$ that is not allowed by observational constraints (see Table 1).
\begin{table}[tbp]
\centering
\begin{tabular}{|p{1in}|p{2in}|}\hline
\multicolumn{2}{|c|}{$N_{eff}=6.77 \thickspace$ and $\thickspace \eta_{BBN} =(4.51\pm0.04)\times10^{-10}$}\\ \hline
$Y_{p}$ & 0.2431 \\ \hline
$D/H\times10^{5}$ & 4.27 \\ \hline
${}^3{He}/H\times {10}^{5}$ &1.23  \\ \hline
${}^7{Li}/H\times {10}^{10}$ & 2.28  \\ \hline
\end{tabular}
\caption{The effect of photon cooling on light element abundances.}
\end{table}
\section{Variation of neutrino temperature}

 During BBN and after the phase of electron-positron annihilation, the ratio of the neutrino temperature to the photon temperature is given by the standard relation \cite{Nollet K. M. and Steigman G. 2015}:
 \begin{dmath}
{\left(\frac{T_{\nu }}{T_{\gamma }}\right)}_{0}^{3}=\frac{4}{2g_s\left(T_{\nu d}\right)-10.5},
\label{Eq.3}
\end{dmath}
where $T_{\nu d}$ is the neutrino decoupling temperature and  $g_s$ is the number of relativistic degrees of freedom contributing to the total entropy, or equivalently, the ratio of the total entropy to the entropy of photons. This determines the effective number of neutrinos given by:
\begin{equation}
N^0_{eff}=3{\left[{\frac{11}{4}\left(\frac{T_{\nu }}{T_{\gamma}}\right)}^3_0\right]}^{4/3}=3{\left[\frac{11}{2g_s\left(T_{\nu d}\right)-10.5}\right]}^{4/3},
\label{Eq.4}
\end{equation}
 These degrees of freedom contribute to the total entropy so that $g_s(T)=\frac{7}{8}\times \left(g_{\nu }+g_{e^{\pm }}\right)+g_{\gamma }=10.75$ where $g_{\nu }=6$, $g_{e^{\pm }}=4$ and $g_{\gamma }=2$ are the degrees of freedom of neutrinos, electron-positron and photon respectively.\\
\noindent The presence of weakly interacting massive particles (WIMP's) will modify Eq.~(\ref{Eq.3}) and Eq.~(\ref{Eq.4}) above. When a WIMP couples to neutrinos or to photons, they speed up the expansion rate because they can alter the energy density of relativistic particles.  Also, the annihilation of WIMP's can heat the photons or the standard neutrino temperature in addition to the heating from $e^{\pm }$ annihilation. In other words, when WIMP's couple to neutrinos they share some of the entropy which causes the neutrino temperature to be different than that of the SBBN.\\
  Therefore, the neutrino temperature can receive different contributions either directly by WIMP interactions or by adding a dark component namely heating /cooling of photon relative to neutrino temperature which will be discussed in section 8. In the next section we represent the neutrino temperature variation by multiplicative factor $\alpha$ regardless of the source.\\

\section{Effect of varying neutrino temperature and chemical potential on the lithium production}
As noted in \cite{Galvez and Sherrer 2017}, since the total neutrino energy density is proportional to $N_{\nu }T^4_{\nu }$ , a variation in the neutrino temperature can be translated into a variation in the effective number of neutrinos through:
\begin{equation}
N_{eff}T^4_{\nu SM}=N_{\nu }T^4_{\nu},
\label{Eq.11}
\end{equation}
where $T_{\nu SM}$ is the standard neutrino temperature, $N_{eff}$ is the effective number of neutrinos derived from the Cosmic microwave Background (CMB) and $N_{\nu }\ $is the number of neutrinos from SBBN. In SBBN $N_{eff}=N_{\nu }$ because $T_{\nu SM}=T_{\nu}$ in Eq.~(\ref{Eq.11}).  However, if we deviate from SBBN, this leads to a difference between $N_{eff}$ and$\ N_{\nu }$ after neutrinos decoupling.
\noindent It is clear from Eq.~(\ref{Eq.11}) that we can vary two parameters while the third one will depend upon the other two. For example, if the neutrino temperature is modified by a factor alpha  $T_{\nu }=\alpha T_{\nu SM}$  for a fixed $N_{\nu }$, then  $N_{eff}$ will depend on these two parameters. As obtained by \cite{Galvez and Sherrer 2017}, $N_{\nu }=5\ $is ruled out for any change in $T_{\nu }$ and for $N_{eff}=3.15\pm 0.23$. In addition, no ranges for $N_{\nu }\ $and $T_{\nu }$ are available that can reduce the lithium abundance significantly. However, the range of $N_{eff}$ is model dependent, and if one relaxes the BBN constraint, one can allow for a broader range based upon the Planck constraint.
In the present work, Tables 2 and 3 are obtained as follows:\\
The neutrino temperature is modified by a factor $\alpha$ where neutrino chemical potentials ${\beta }_{{\nu }_{\mu ,\tau ,e}}$ are taken to be equal. Then the total neutrinos energy density is given by:
 \begin{equation}
 {\rho }_{\nu }+{\rho }_{\overline{\nu }}=\frac{7{\pi }^2}{120}gT^4_{\nu }\left[1+\frac{{30\beta }^2}{7{\pi }^2}+\frac{{15\beta }^4}{7{\pi }^4}\right]
\label{Eq.13}
\end{equation}
Using Eq. (~\ref{Eq.13}), Eq. (~\ref{Eq.11}) is modified as follows:
\begin{dmath}
N_{eff}T^4_{\nu SM}=N_{\nu 0 }\left(1+\frac{{30\beta }^2}{7{\pi }^2}+\frac{{15\beta }^4}{7{\pi }^4}\right)T^4_{\nu }+{\Delta N}_{\nu }T^4_{\nu }
\label{Eq.14}
\end{dmath}
Eq. (~\ref{Eq.14}) includes the three standard neutrino species $N_{\nu 0}$ with their chemical potentials while ${\Delta N}_{\nu }T^4_{\nu }$ is for extra relativistic species not contributing to the chemical potential with ${\Delta N}_{\nu } = N_{\nu}-N_{\nu 0}$.
Restricting ourselves to the predicted light element abundances by SBBN we obtain the following range of $N_{\nu }$,  $N_{eff}$ and ${\beta }_{{\nu }_{\mu ,\tau ,e}}$ which reproduces the SBBN predictions:
\begin{equation}
\begin{split}
3\le N_{\nu }\le 20, \thickspace  -0.01 \le {\beta }_{{\nu }_{\mu ,\tau ,e}}\le  0.32,\\ 2.27 \le N_{eff}\le  3.25 , \thickspace  0.58\le \alpha \le 1.02,
\label{Eq.15}
\end{split}
\end{equation}
where $N_{eff}$ is compatible with that deduced by \cite{Plank Collaboration 2016} and \cite{WMAP2013}. The inspection of Table 2 is to show that the abundances predicted by SBBN can be obtained by this non-standard assumption described above. In other words, despite of these variations of neutrino temperature and chemical potentials, the abundances predicted by SBBN are not affected.
\begin{table}[tbp]
\centering
\begin{tabular}{|p{0.6in}|p{0.4in}|p{0.5in}|p{0.4in}|p{0.6in}|p{0.8in}|p{1in}|} \hline
${\beta }_{{\nu }_{\mu ,\tau ,e}}$ & $\alpha$ & $N_{eff}$ & $N_{\nu }$ & $Yp$ & $D/H\times {10}^{5}$ & ${}^7{Li}/H\times {10}^{10}$ \\ \hline
0.00 & 1.00 & 3.0000 & 3 & 0.2461  & 2.6386 & 4.3861 \\ \hline
0.10 & 0.88 & 3.0063 & 5 & 0.2423 & 2.6219 & 4.3277 \\ \hline
0.14 & 0.85 & 3.1454 & 6 & 0.2401 & 2.6466 & 4.2310 \\ \hline
0.2 & 0.77 & 3.1821 & 9 & 0.2413 & 2.6623 & 4.2260 \\ \hline
\end{tabular}
\caption{Light elements abundances with varying neutrino number, chemical potential and temperature.}
\end{table}
However, in order to obtain a substantial reduction of the lithium abundance, the ranges of Table 2 must be extended. This is achieved in Table 3 if the following ranges are adopted:

\begin{equation}
\begin{split}
3\le N_{\nu }\le 20, \thickspace 0.06\le {\beta }_{\nu_{\mu ,\tau ,e}}\le 0.43,\\  6.67\le N_{eff}\le 7.79,\thickspace 0.76\le \alpha \le 1.27
\label{Eq.16}
\end{split}
\end{equation}
 \begin{table}[tbp]
\centering
\begin{tabular}{|p{0.6in}|p{0.4in}|p{0.5in}|p{0.4in}|p{0.6in}|p{0.8in}|p{1in}|} \hline
${\beta }_{{\nu }_{\mu ,\tau ,e}}$ & $\alpha$ & $N_{eff}$ & $N_{\nu }$ & $Yp$ & $D/H\times {10}^{5}$ & ${}^7{Li}/H\times {10}^{10}$\textbf{} \\ \hline
0.06 & 1.23 & 6.8773 & 3 & 0.2337 & 3.5492 & 2.7993 \\ \hline
0.13 & 1.16 & 7.2825 & 4 & 0.2299 & 3.6124 & 2.6764 \\ \hline
0.12 & 1.1 & 7.3480 & 5 & 0.2426& 3.7261 & 2.7538 \\ \hline
0.24 & 0.98 & 7.4484 & 8 & 0.2351 & 3.6876 & 2.6807 \\ \hline
\end{tabular}
\caption{Effect of varying neutrino number, chemical potential and temperature on lithium.}
\end{table}
We can conclude the following: if we choose any  $N_{\nu }$ we can achieve a reduction of lithium by a combination of $\alpha$ and ${\beta }_{{\nu }_{\mu ,\tau ,e}}$. This illustrates the importance of these two parameters.
We emphasize that the above results are compatible with all successful models that lead to a substantial decrease in lithium at the expense of increasing deuterium to the maximum value allowed by observations as stated in Eq.~(\ref{Eq.0}). Our variation of $N_{\nu }$ as above tends to affect directly the expansion rate, while the effect of varying the neutrino temperature is more complicated:
 \begin{enumerate}
\item This factor $\alpha$ will alter the energy density of neutrinos (see Eq. (~\ref{Eq.13})) and consequently the expansion rate given by Friedmann equation.
\item Multiplying $T_{\nu}$ by a factor $\alpha$ will modify the effective relativistic degrees of freedom but the effect on neutron mass fraction at the freeze-out is not significant. In other words, since variation of neutrino temperature is taken after neutrino decoupling (after the freeze-out of neutrons), neutron mass fraction at the freeze-out is not significantly affected. Meanwhile, this variation of neutrino temperature will affect deuterium bottleneck temperature/time and consequently final deuterium abundance. Helium is slightly affected due to its dependance also on the temperature when deuterium reaches its bottleneck. This is shown by some analytical calculations following the ones given by \cite{Mukhanov V. 2005} (see \cite{Makki2018} for detailed discussion). Those analytical expressions allow us to understand the dependance of light elements into three stages of BBN namely, the freeze-out of neutrons, deuterium bottleneck and when neutron concentration drops to that of deuterium.
\end{enumerate}
This successful calculation to decrease lithium was obtained with the range of $N_{eff}$ shown above, however, this is not supported by recent CMB observations as discussed in section 2. To achieve a reduction of lithium without violating CMB limits on $N_{eff}$, one possible way is to add the effect of photon cooling which we describe in the next section.

\section{Effect of photon cooling}

In this section, we add effect of photon cooling with axion to the previous variations done before for the aim to decrease lithium abundance without violating observational constraints on $N_{eff}$ and other light elements. The baryon to photon ratio as determined by Planck \cite{Plank Collaboration 2016} is given to be $\eta =(6.11\pm0.06)\times10^{-10}$, however, photon cooling through axions after BBN implies that $\eta_{BBN} =(4.51\pm0.04)\times10^{-10}$ (see Eq. (~\ref{Eq.2})). Photon cooling with axions will also modify the effective degrees of freedom. The energy density of the universe can be written as:
\begin{equation}
\rho_{rad}=\rho_{\gamma}[1+N_{eff}\times\frac{7}{8}\times(\frac{4}{11})^{\frac{4}{3}}]
\label{Eq.18}
\end{equation}
On the other hand, the radiation density after BBN is given to be:
\begin{equation}
\rho_{rad}=\rho_{\gamma}[1+\frac{1}{2}+N_{\nu}\times\frac{7}{8}\times(\frac{4}{11})^{\frac{4}{3}}\times\frac{3}{2}]
\label{Eq.19}
\end{equation}
where the factor 3/2 is due to the photon cooling relative to neutrinos and 1/2 represents axion degree of freedom. In this case the relativistic degrees of freedom observed now is given to be
\begin{equation}
N_{eff}=\frac{3}{2}N_{\nu}+\frac{1}{2}\times\frac{8}{7}\times(\frac{11}{4})^{\frac{4}{3}}
\label{Eq.20}
\end{equation}
If we take such case which is treated by \cite{Erken2012}, light elements abundances are given in Table 1 where the lithium decreases to match observations but this will increase deuterium and $N_{eff}$ to values that are not supported by observations.\\
In this paper, in addition to photon cooling, we vary neutrinos temperature in order to conserve the constraints on $N_{eff}$ and deuterium. In addition, we vary chemical potential to obtain additional decrease in lithium without violating observational constraints on helium. Introducing photon cooling, varying neutrino temperature and chemical potential, we modify the radiation energy density given in Eq. (~\ref{Eq.19}) as follows:
\begin{equation}
\rho_{rad}=\rho_{\gamma}[1+\frac{1}{2}+\frac{7}{8}\times(\frac{4}{11})^{\frac{4}{3}}\times\frac{3}{2}\times\alpha^{4}(\Delta N_{\nu}+N_{\nu0}(1+\frac{{30\beta }^2}{7{\pi }^2}+\frac{{15\beta }^4}{7{\pi }^4}))]
\label{Eq.21}
\end{equation}
Then Eq. (~\ref{Eq.21}) will lead to the new effective degrees of freedom,
\begin{equation}
N_{eff}=\frac{1}{2}\times\frac{8}{7}\times(\frac{11}{4})^{\frac{4}{3}}+\frac{3}{2}\alpha^{4}[\Delta N_{\nu}+N_{\nu0}(1+\frac{{30\beta }^2}{7{\pi }^2}+\frac{{15\beta }^4}{7{\pi }^4})]
\label{Eq.22}
\end{equation}
It clear from Eq. (~\ref{Eq.22}) that multiplying neutrinos temperature with a factor $\alpha<1$ will decrease the relativistic degrees of freedom to match observations discussed in Section 3. In addition, it will decrease deuterium so that the overabundance observed in Table 1 can be reduced. Restricting the relativistic effective degrees of freedom $2<N_{eff}<5$ we performed numerical simulations after updating the code given by \citet{Arbey A. 2012}  to obtain the ranges shown in Eq. (~\ref{Eq.23}) and Table 4. It is seen from Table 4 that $N_{eff}$ is now compatible with recent CMB data but still a bit higher than Planck one. In addition, the lithium decreases significantly at the expenses of increasing deuterium but to a value that is allowed by observations.
\begin{equation}
\begin{split}
3\le N_{\nu }\le 20, \thickspace 0\le {\beta }_{\nu_{\mu ,\tau ,e}}\le 0.34,\\
4.3108\le N_{eff}\le 4.9609,\thickspace 0.52\le \alpha \le 0.88
\label{Eq.23}
\end{split}
\end{equation}
It is important to mention here that deuterium observations and predictions still matter of debate. So far, there are 15 measurements of the primordial deuterium values that show a large dispersion in the mean values and in the estimated errors \cite{Balashev S. A. 2016}.
Meanwhile, if we want to take the most precise measurement of these 15 values by \cite{Cooke 2014}, $\frac{D}{H}=(2.53\pm0.04)\times10^{-5}$, there is still still a tension between such measurements and BBN calculations where  $\frac{D}{H}=(2.65\pm0.07)\times10^{-5}$ \cite{Valentino E. Di} or the standard value of $\frac{D}{H}$ obtained in this work. Measurements of nuclear cross sections for deuterium production and destruction are still under revisions in order to reduce errors coming from nuclear cross section and to remove the tension between predictions and observations. Therefore, taking the range of observations given in  Eq. (~\ref{Eq.0}) is conceivable, first because it is early to talk about deuterium plateau due to the big dispersion in the observations. Second, deuterium is easily destroyed than ${}^7$Li so that higher deuterium values may be more representative of its primordial one \cite{Olive K. et al. 2012} and the dispersion in the observations is due to unknown stellar processes \cite{Erken2012}.
 \begin{table}[tbp]
\centering
\begin{tabular}{|p{0.6in}|p{0.4in}|p{0.5in}|p{0.4in}|p{0.6in}|p{0.8in}|p{1in}|} \hline
${\beta }_{{\nu }_{\mu ,\tau ,e}}$ & $\alpha$ & $N_{eff}$ & $N_{\nu }$ & $Yp$ & $D/H\times {10}^{5}$ & ${}^7{Li}/H\times {10}^{10}$\textbf{} \\ \hline
0.05 & 0.83 & 4.3372& 3 &0.2313 & 3.4847 & 2.7977 \\ \hline
0.11 & 0.77 & 4.8381 & 5 & 0.2342 & 3.6563 & 2.6684 \\ \hline
0.16 & 0.7 & 4.7227 & 7 & 0.2335& 3.6165 & 2.6963 \\ \hline
0.25 & 0.58 & 4.9176 & 16 & 0.2402 & 3.7264 & 2.6890 \\ \hline
\end{tabular}
\caption{Effect of adding photon cooling on lithium along with non-standard neutrino properties.}
\end{table}
\section{The effect of a unified dark fluid}
We have seen in section 8 that adding the axion dark matter was successful to decrease the ${}^7$Li abundance with $N_{eff}$ compatible with the recent CMB observations, but still higher than the one determined by Planck mission. In order to satisfy the requirements on $N_{eff}$ we found a way by adopting the so called dark fluid along with non-standard neutrino properties. We know that dark matter may be modeled as a system of collision-less particles while dark energy may be described as a scalar field in the context of quintessence model \cite{Arbey A. 2009}. However, in this section a unified fluid \cite{Arbey A. 2005} is adopted to describe the dark energy and dark matter as two different aspects of the same component.
 To explore this scenario, a temperature-dependent dark energy density can be added to the radiation density as follows \cite{Arbey A. 2009}:
\begin{dmath}
{\rho }_D\left(T\right)=k_{\rho}{\rho}_{rad} \left(T_0\right){\left(\frac{T}{T_0}\right)}^{n_{\rho }}
\label{Eq.6}
\end{dmath}
where $\rho_{rad}=\rho_{\gamma}+ \rho_{e^{\pm}}+\rho_{\nu}$, $T_0$=1.0 Mev $=1.16\times {10}^{10}$K , $k_{\rho}$ is the ratio of effective dark fluid density over the total radiation density at $T_{0}$ and $n_{\rho }$ characterizes the adiabatic expansion of the fluid. In the case of $n_{\rho }=4$, the dark component mimics a radiation density. The case $n_{\rho }=3$ describes a matter density, while $n_{\rho }=6$ can describe a scalar field. With these assumptions the Friedmann equation during BBN should be modified as:
\begin{equation}
{\left(\frac{\dot{a}}{a}\right)}^2=H^2=\frac{8\pi G}{3}\left({\rho }_{rad}+{\rho }_D\right)
\label{Eq.7}
\end{equation}
In analogy, the temperature-dependent dark entropy can be added as follows:
\begin{dmath}
{\ s}_D\left(T\right)=k_ss_{rad}\left(T_0\right){\left(\frac{T}{T_0}\right)}^{n_s}
\label{Eq.8}
\end{dmath}
where
\begin{dmath}
s_{rad}\left(T\right)=g_{s}\left(T\right)\frac{{2\pi }^2}{45}T^3
\label{Eq.9}
\end{dmath}
and  $g_{s}\left(T\right)$ is the effective number of degrees of freedom characterizing the contribution of relativistic particles to the entropy density.
Then, the total entropy becomes:
\begin{equation}
s_{tot}\left(T\right)=s_{rad}\left(T\right)+s_D\left(T\right)
\label{Eq.10}
\end{equation}
It is important to emphasize that the dominant effects of adding dark component described above is to alter the expansion rate, the time-temperature relation, the neutrino-to-photon temperature, and consequently the light element abundances.\\
It is clear from Eqs. (~\ref{Eq.6},~\ref{Eq.8}) that including this dark component, four parameters are introduced $k_{\rho}, n_{\rho}, k_{s}, n_{s}$. Knowing that the universe was radiation dominated during the time of BBN requires the constraints $ n_{\rho}\geq4$ and $ k_{\rho}<1.0$ \cite{Arbey A.and Mahmoudi F. 2008}. Then, for fixed values of $k_{\rho}, n_{\rho}, k_{s}, n_{s}$ we will vary neutrino number, chemical potentials and temperature in order to decrease lithium without violating observational constraints on light elements and $N_{eff}$. Our calculations summarized in Table 5 shows the following:\\
Fixing $k_{\rho}=0.12$, $ n_{\rho}=6$, $k_{s}=0.00045$ and $n_{s}=5$, we are able to find for every chosen value of $N_{\nu}$ a combination of $\alpha$ and $\beta_{{\nu}_{e,\mu,\tau}}$ that reduces lithium and conserves the constraints on $N_{eff}$. Note that $N_{eff}$ is still given by Eq. (~\ref{Eq.14}) because the dark component is effective during BBN due to the power law (see Eqs. (~\ref{Eq.6},~\ref{Eq.8})) and have negligible contribution after BBN.
Eq. (~\ref{Eq.24}) shows the ranges of $N_{\nu}$, $\alpha$, and $\beta_{{\nu}_{e,\mu,\tau}}$ that lead to a decrease of lithium and we see now with this treatment that $N_{eff}$ is compatible with recent CMB data including Planck and WMAP ones.
\begin{equation}
\begin{split}
k_{\rho}=0.12, \thickspace  n_{\rho}=6, \thickspace  k_{s}=0.00045 \thickspace, n_{s}=5,\\
3\le N_{\nu }\le 20,\thickspace 0.07\le {\beta }_{\nu_{\mu ,\tau ,e}}\le 0.45, \\
2.9553\le N_{eff}\le 3.8486, \thickspace 0.62\le \alpha \le 1.03
\label{Eq.24}
\end{split}
\end{equation}
Finally, we can choose other values of dark component parameters so other ranges of $\alpha$, $\beta_{{\nu}_{e,\mu,\tau}}$ and $N_{eff}$ can be derived, but for simplicity we have fixed dark component parameters in order to give a better understanding of the physical conditions under which the BBN has been operating and provide many possibilities to ameliorate the lithium problem.
\begin{table}[tbp]
\centering
\begin{tabular}{|p{0.6in}|p{0.4in}|p{0.5in}|p{0.4in}|p{0.6in}|p{0.8in}|p{1in}|} \hline
\multicolumn{7}{|c|}{$k_{\rho}=0.12$, $ n_{\rho}=6$, $k_{s}=0.00045$ and $n_{s}=5$}\\ \hline
${\beta }_{{\nu }_{\mu ,\tau ,e}}$ & $\alpha$ & $N_{eff}$ & $N_{\nu }$ & $Yp$ & $D/H\times {10}^{5}$ & ${}^7{Li}/H\times {10}^{10}$\textbf{} \\ \hline
0.14 & 1.03 & 3.4053& 3 &0.2310 & 3.5892 & 2.6890 \\ \hline
0.2 & 0.93 & 3.0313 & 4 & 0.2291 & 3.4669 & 2.7847 \\ \hline
0.22 & 0.89 & 3.1768 & 5 & 0.2319& 3.5281 & 2.7639 \\ \hline
0.34 & 0.76 & 3.6698 & 11 & 0.2333 & 3.6912 & 2.6249 \\ \hline
\end{tabular}
\caption{Effect of unified dark fluid on lithium along with non-standard neutrino properties.}
\end{table}
\section{Conclusion}
In order to reduce the primordial ${}^7$Li abundance by Standard Big Bang Nucleosynthesis (SBBN) without violating constraints on the abundances of light elements and on the effective number of relativistic degrees of freedom $N_{eff}$ from CMB observations (see section 3), we firstly considered the effects of the variation of neutrino number, chemical potential and temperature. Secondly, it was necessary to add the effect of dark components. Our results may be summarized as follows:

\begin{itemize}
\item It was possible to reproduce SBBN predictions for a certain range of neutrino temperature, number and chemical potential (see section 7 and Table 2). This finding motivated a non-standard treatment that may lead to a reduction of the lithium abundance.
\item Extending the range of variation of the above parameters was successful to reduce the lithium abundance, but with $N_{eff}$ not compatible with recent analysis of the CMB. One way  out was to add photon cooling as described in section 8, which led to a reduction of lithium abundance with  $N_{eff}$ compatible with recent CMB observations discussed in section 3. However, this range of $N_{eff}$ is still in tension with the most precise one deduced by Planck mission.
 \item Taking into considerations a unified dark fluid along with non-standard neutrino properties has led to a decrease in lithium for ranges of  $N_{eff}$ that satisfies recent CMB observations including Planck and WMAP ones.
\item As a final remark, we think that decreasing the primordial lithium abundance below the predictions of the SBBN is achieved only with a maximum deuterium abundance which is still justified by observations (see \cite {Balashev S. A. 2016}). However, it is worth investigating the possibility of reducing lithium abundance avoiding the increase of deuterium by including new particles or interactions \cite{Goudelis 2016}, \cite{Salvati 2016}. The lithium problem is really challenging.

\end{itemize}

\end{document}